\documentclass[a4paper,english,amsmath,amssymb,superscriptaddress,12pt]{revtex4}

\usepackage[latin1]{inputenc}
\usepackage[T1]{fontenc}
\usepackage{babel}

\usepackage{graphicx}
\usepackage{psfrag}
\usepackage{array}

\begin{document}

\title{Computing stationary distributions in equilibrium and non-equilibrium systems with Forward Flux Sampling}

\author{Chantal Valeriani}
\address{FOM Institute for Atomic and Molecular Physics, Kruislaan 407, 1098 SJ Amsterdam, The Netherlands.}
\author{Rosalind J. Allen}
\address{SUPA, School of Physics, The University of Edinburgh, Mayfield Road, Edinburgh EH9 3JZ, UK.}
\address{FOM Institute for Atomic and Molecular Physics, Kruislaan 407, 1098 SJ Amsterdam, The Netherlands.}
\author{Marco J. Morelli}
\address{FOM Institute for Atomic and Molecular Physics,Kruislaan 407, 1098 SJ Amsterdam, The Netherlands.}
\author{Daan Frenkel}
\address{FOM Institute for Atomic and Molecular Physics,Kruislaan 407, 1098 SJ Amsterdam, The Netherlands.}
\author{Pieter Rein ten Wolde}
\address{FOM Institute for Atomic and Molecular Physics,Kruislaan 407, 1098 SJ Amsterdam, The Netherlands.}

\begin{abstract}
  We present a method for computing stationary distributions for
  activated processes in equilibrium and non-equilibrium systems using
  Forward Flux Sampling (FFS). In this method, the stationary distributions are obtained directly
  from the rate constant calculations for the forward and backward
  reactions; there is no need to perform separate calculations for the stationary distribution and
  the rate constant. We apply the method to the non-equilibrium rare event problem
  proposed by Maier and Stein, to nucleation in a 2-dimensional Ising
  system, and to the flipping of a genetic switch.
\end{abstract}

\maketitle

\section{Introduction}
Rare events are ubiquitous in physics, chemistry, and biology;
examples include crystal nucleation, chemical reactions, and protein
folding. Rare events are activated processes, for which the average
waiting time between events can be orders of magnitude longer than the
duration of the event itself. This makes these events intrinsically
difficult to investigate experimentally. Computer simulations are
therefore a natural tool to use - yet conventional numerical
techniques are impractical for rare events, because most of the CPU
time is wasted on the uneventful waiting time between events. A number
of ``rare event'' simulation schemes have recently been developed in
the field of soft-condensed matter physics, which make it possible to
zoom in on the rare events themselves. Techniques such as umbrella
sampling allow the calculation of free-energy barriers separating the
stable states \cite{umbrella1,umbrella2,umbrella3,umbrella4}, while
schemes such the Bennet-Chandler method \cite{bencha1,bencha2} also allow the 
computation of rate constants. Transition path sampling \cite{dellago1,dellago2,vanerp1,vanerp2} allows both rate constants and transition paths to be obtained. These techniques have been
used for a wide range of applications including ion permeation through
membranes, protein folding, and nucleation. However, these schemes
require prior knowledge of the phase-space density. For systems that
are in thermodynamic equilibrium---with detailed balance and
microscopic reversibility---the phase-space density is known: it is
given by the Boltzmann distribution. In contrast, for systems that are
out of equilibrium, the phase-space density is usually not known. This
means that most numerical techniques for simulating rare events are
limited to equilibrium systems, and thus exclude a host of important
rare-event problems in non-equilibrium systems, such as polymer
collapse under flow, crystal nucleation under shear, and rare events
in biology, such as protein translocation and switching events in
biochemical networks. We have recently developed a numerical
technique, called Forward Flux Sampling (FFS)\cite{FFS,FFS2,FFS3}, that makes it possible
to compute rate constants in both equilibrium and non-equilibrium
systems with stochastic dynamics. In this paper, we show how
stationary distributions can also be obtained directly from an FFS
calculation, for both equilibrium and non-equilibrium systems. For
equilibrium systems the advantage is that from an FFS simulation one
can obtain not only the rate constant, but also information about the free-energy
landscape. For non-equilibrium systems the concept of free energy does
not apply, but one can obtain the steady-state probability
distribution as a function of a chosen order parameter (or order parameters). To our
knowledge, this is the first method to be proposed for computing
stationary distributions for multi-dimensional non-equilibrium systems
that are in steady state.

In soft-condensed matter physics, the rate $k$ of an activated process in
an equilibrium system is often written as the product of two
factors:
\begin{equation}
\label{eq:k_RP}
k = R (q^*) \rho(q^*).
\end{equation}
Here, $q$ is an order parameter that connects the initial and final
states, assuming that the system evolves between two states. It is defined such that for $q<q^*$, the system is in the
initial state, while for $q>q^*$ it is in the final state. The
quantity $\rho(q^*)$ is the probability that the system is at the
dividing surface $q=q^*$ and $R (q^*)$ is the rate at which this
dividing surface is crossed. For equilibrium systems, $\rho(q^*)$ is
proportional to $\exp(-\beta \Delta G(q^*))$, where $\beta$ is the
inverse temperature and $\Delta G(q)$ is the (Landau) free energy of
the system as a function of the order parameter $q$.  It is natural to
locate the dividing surface $q^*$ at the top of the free-energy
barrier $\Delta G(q)$ separating the two states. The
rate constant is thus given by the probability of being at the top of the
free-energy barrier, multiplied by a kinetic prefactor.

 The Bennett-Chandler method for computing rate constants for activated processes uses a two-step procedure \cite{bencha1,bencha2}: one  first computes the free-energy
barrier, using, for example, the umbrella sampling scheme
\cite{umbrella1,umbrella2,umbrella3,umbrella4}, and then  the kinetic prefactor, using a Molecular Dynamics simulation in
which trajectories are fired from the top of the free-energy barrier. However, this method is
computationally demanding, and its success depends strongly on
the choice of the reaction co-ordinate $q$. If $q$ is poorly chosen, the
system will sample the wrong part of the phase space, which will not
only conceal the mechanism of the transition, but also impede the
computation of the rate constant---while the choice of $q$ co-ordinate does not affect the value of the rate constant $k$, it can strongly affect  the efficiency with which $k$ is
computed. For high-dimensional complex
systems it can be difficult to make a good choice for $q$, since this
requires {\em a priori} insight into the reaction mechanism.

Transition-path sampling (TPS) has been developed to alleviate these problems \cite{Pratt,dellago1,dellago2,Bolhuis}. This scheme generates an
ensemble of trajectories between the initial and final states using Monte Carlo sampling in trajectory space. TPS only
requires an order parameter to distinguish the initial and final
states; this order parameter does not need to be the true
reaction co-ordinate. TPS thus makes it possible to compute the rate
constant without prior knowledge of the reaction mechanism. However,
this method does require knowledge of the steady-state phase space distribution, which is needed for the acceptance/rejection step in the Monte-Carlo scheme, and it does not allow direct computation of the free-energy barrier
separating the two states. Moroni {\em{et al.}}  have developed a related
method, transition interface sampling (TIS), which relies on
the computation of crossing probabilities of a series of interfaces
between the initial and final states \cite{Moroni1,Moroni2,vanerp1,vanerp2}. A variant of this method, partial path TIS (PPTIS),
assumes loss of time correlations in the transition paths over a
distance of two interfaces. Moroni {\em{et al.}}
have recently shown how the free-energy barrier as well as the rate constant can be obtained from a single TIS / PPTIS calculation
\cite{Moroni3}. As for TPS, both TIS and PPTIS require the system to be in thermodynamic equilibrium. The
``milestoning'' method of Faradjian and Elber also employs a series
of interfaces to compute rate constants and also assumes that the
interface-crossing probability does not depend upon the full history
of the path \cite{Faradjian04}. In an alternative approach, Vanden-Eijnden {\em{et al}} have developed a set of ``string'' methods, which can be used to compute minimum free-energy paths and
the probability current of reactive trajectories for equilibrium systems
\cite{Ren02,Maragliano06}. 

The algorithms discussed above---TPS, (PP)TIS, milestoning and the
string methods --- are limited to 
systems that are in thermodynamic equilibrium. The Forward Flux Sampling (FFS) method, and its variants,  were developed to calculate rate constants and transition paths for rare events in  equilibrium and non-equilibrium systems with
stochastic dynamics \cite{FFS,FFS2,FFS3}. Like TIS, PPTIS, and milestoning, FFS uses a
series of interfaces to compute the rate constant. However (unlike PPTIS and milestoning), 
FFS does not make the Markovian assumption that
the distribution of paths at the interfaces is independent of the
path histories. The order parameter that is used to
define the location of the interfaces need not be the reaction
co-ordinate, and the choice of order parameter, in principle, does not bias the dynamics of the transition paths.

We have recently shown that the rate constant for activated processes in
non-equilibrium systems that are in steady state can also be written in the form of Eq.~\ref{eq:k_RP} \cite{Warren05}. The
quantity $\rho(q)$ is then the stationary probability distribution
function for the order parameter $q$. In this paper we show that the stationary distribution $\rho(q)$, as well as the forward and backward rate
constants and transition paths, can be obtained by
performing two FFS calculations---one for the transition from the initial to
the final state, and  the other for the reverse transition. 
The method can be applied to both non-equilibrium and equilibrium systems; in the latter case,
$\rho(q)$ corresponds to the Boltzmann distribution. The method is
conceptually similar to that used in TIS and PPTIS to compute
free-energy barriers, in the sense that the stationary distribution
$\rho(q)$ is obtained by matching the forward and backward trajectories
\cite{Moroni1,vanerp1,vanerp2,Moroni2}.

In the next section, we explain the FFS algorithm \cite{FFS}
. In sections \ref{sec:st} and \ref{sec:sm}, we discuss the theory and method for obtaining stationary distributions. We then
illustrate the method using symmetric and asymmetric double-well
potentials (section \ref{sec:oned}), and the two-dimensional  non-equilibrium rare event problem proposed by Maier and Stein (section \ref{sec:ms}). In section
\ref{sec:ising}, we use the method to calculate the free-energy barrier for
nucleation in a two dimensional Ising system. Finally, in section \ref{sec:gs}, we compute non-equilibrium stationary probability distributions for  a bistable model genetic switch.

\section{Forward Flux Sampling}
\label{sec:FFS}

We consider rare, spontaneous transitions between two regions of state
space $A$ and $B$. The phase space co-ordinates are denoted by $x$ and
the regions $A$ and $B$ are defined in terms of an order parameter
$\lambda(x)$ such that the system is in
state $A$ if $\lambda(x) < \lambda_0$, and it is in state $B$ if $\lambda(x) > \lambda_n$. The key principle is to use a series of
interfaces $\lambda_0, \lambda_1, \dots, \lambda_{n-1}, \lambda_n$, to
drive the system from state $A$ to state $B$ in a ratchet-like
manner. The idea of the interfaces is that  
they make it possible to capitalize on all those fluctuations that  
bring the system in the direction of the final state B. 

Supposing that with a conventional (say MD) simulation, the system exhibits a rare  
fluctuation that moves it up the barrier, and that it  
crosses an interface between state A and the top of the barrier, 
if we would continue this succesfull run, then most likely the system would roll back  
down the hill, i.e.  relax back towards state A, and one would have  
to wait for "another" rare fluctuation that moves the system in the  
direction of B. By storing the configurations at the interfaces, we   
can thus efficiently exploit all those fluctuations that move the  
system up the barrier.

In this paper, we make use of 
the original FFS scheme~\cite{FFS} presented in more details in Ref.~\cite{FFS2}.

In FFS, one first performs a conventional, brute-force
simulation in state $A$. Each time the
system crosses the interface $\lambda_0$ in the direction of increasing $\lambda$ during this simulation, the co-ordinates of that state
point are stored. One also measures the average number per unit time of these crossings. At the end of this simulation, one has a measure of the flux $\Phi_A$ of trajectories crossing $\lambda_0$ from $A$, as well as a 
collection of state points corresponding to crossings of the first interface, $\lambda_0$, coming from $A$. This
collection is then used to provide starting points for a set of trajectories, each of which is continued until the system either
reaches the next interface, $\lambda_1$, or returns to state $A$ ({\em{i.e.}} reaches $\lambda_0$). This procedure
generates a new collection of state points at the next interface, which are the end points of those trajectories that arrived at $\lambda_1$ from $\lambda_0$. One also obtains an estimate of the probability $P(\lambda_{1}|\lambda_{0})$ that a trajectory which reaches $\lambda_0$ from $A$ will subsequently reach $\lambda_1$ without returning to $A$ - this is simply the fraction of trajectories which arrive at $\lambda_1$.
By repeating this procedure for all subsequent interfaces, 
one has for each interface $i$ an 
estimate of the probability $P(\lambda_{i+1}|\lambda_{i})$ that, given  that a trajectory has reached interface $i$ coming from $A$, it subsequently reaches $\lambda_{i+1}$ before returning to
$A$. The rate constant $k_{AB}$ can then be obtained from \cite{Moroni3}
\begin{equation}
k_{AB} = \Phi_A \prod_{i=0}^{n-1} P(\lambda_{i+1}|\lambda_{i}).
\label{eq:kAB}
\end{equation}
By tracing back paths that successfully arrive at $\lambda_n$, one can also sample the transition path ensemble for the rare event. Analysis of these paths can lead to insight into the mechanism by which the event occurs.

\section{Stationary distributions: theory}
\label{sec:st}
We are interested in computing the stationary distribution
$\rho(q)$, where $\rho(q)dq$ is the probability of observing the order parameter $q$ in the range $q \to q+dq$, for a system that is in a stationary state. 
We stress the fact that the order parameter $q$ for the
computation of the stationary distribution function need not be the
same as the order parameter $\lambda$ that is chosen for the FFS
calculation. 
The stationary distribution can be expressed as   
\begin{equation}
\rho(q) = \langle\delta(q-q(x))\rangle,
\label{eq:rho_q}
\end{equation}  
where $x$ is a point in the multi-dimensional phase space. For
equilibrium systems, the contributions to the average in Eq.~\ref{eq:rho_q} are weighted
according to the Boltzmann distribution, while for non-equilibrium
systems they are weighted according to the
steady-state phase-space density.  For both equilibrium and
non-equilibrium systems that are in steady state and ergodic, this
ensemble average is equivalent to a time average over a long brute-force simulation, in which $\rho(q)$ measures the frequency with which value $q$ of the order parameter is ``visited'' by the trajectory.

The distribution function $\rho(q)$ is easy to sample close to the stable states $A$ and $B$, using conventional, brute-force simulation. However, this method will lead to poor statistics in the ``barrier'' region between $A$ and $B$, which is rarely visited. We use FFS to obtain $\rho(q)$ in the ``barrier'' region, and supplement this with conventional sampling in the two stable states to obtain the complete distribution function.

\begin{figure}[h!]
\begin{center}
{\rotatebox{0}{{\includegraphics[scale=0.35,clip=true]{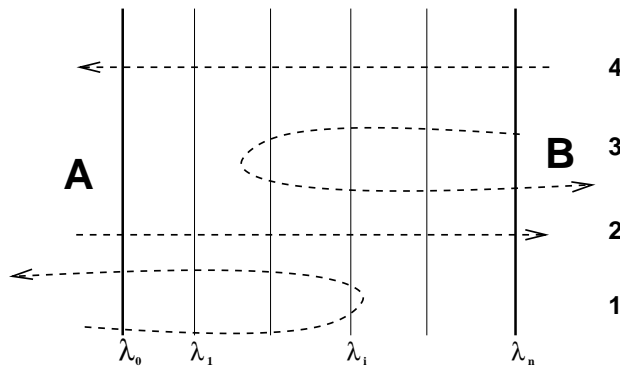}}}}
\caption{A sketch of all the possible trajectories that
contribute to the stationary distribution $\rho(q)$ (see
Eq.~\ref{eq:rho_q}): Trajectory 1 comes from A and goes back to A; trajectory 2 comes from A
  and goes to B; trajectory 3 comes from B and goes back to B;
  trajectory 4 comes from B and goes to A. The FFS simulation
  from $A$ to $B$ harvests the trajectories corresponding to types $1$ and
  $2$, while the FFS simulation for the reverse transition generates
  trajectories of types $3$ and $4$. The interfaces $\{\lambda_0,
    \lambda_1, \dots, \lambda_{n-1}, \lambda_n\}$ used in the FFS
  simulation are also shown. \label{fig:sketch}}
\end{center}
\end{figure}

The key idea which we use to obtain stationary distributions with FFS is to divide the ``visits'' of an imaginary, very long simulation trajectory to  value $q$ of the order parameter into two categories, according to whether the trajectory was most recently in state $A$ or state $B$. We therefore write $\rho(q)$ as the sum of two contributions
\begin{equation}
\rho(q) = \psi_A (q) + \psi_B(q),
\label{eq:rho_phi_A_B_q}
\end{equation}
where $\psi_{A} (q)$ is the contribution to the probability density
$\rho(q)$ from those trajectories that come from region $A$, and $\psi_B(q)$ is the
contribution due to trajectories coming from $B$ (see
Fig.~\ref{fig:sketch}). 

In the basins of attraction $A$ and $B$, the
trajectories will quickly lose memory of where they came from - {\em{i.e.}} we expect excursions out of a basin of attraction to be uncorrelated. This, as
we describe below, makes it possible to obtain the distribution
function $\psi_A(q)$  from an FFS simulation for the transition from $A$
to $B$, while $\psi_B(q)$ can be computed using an FFS simulation for
the reverse transition (see Fig.~\ref{fig:sketch}). For equilibrium
systems, the free energy profile can be obtained from $\Delta G \sim -k_BT
\ln\left[\rho(q)\right]$ once $\rho(q)$ is known.

The function $\psi_A(q)$ is given by
\begin{eqnarray}
\psi_A(q)&=&  p_A \Phi_A \, \tau_+(q;\lambda_0).
\label{eq:phiA_q}
\end{eqnarray}
Here, $p_A$ is the probability that the system is in state $A$ and
$\Phi_A$ is, as in Eq.~\ref{eq:kAB}, the flux of trajectories leaving
state $A$ (i.e. crossing the surface $\lambda_0$ coming from $A$). The quantity
$\tau_+(q;\lambda_0) \equiv \langle\delta(q-q(x))\rangle_{\lambda_0}$ is
the average time spent at order parameter $q$ by a trajectory that originates from interface $\lambda_0$. We note that $\tau_+(q;\lambda_0)$ includes contributions both from paths that start in $A$ and ultimately reach $B$, and from those that start in $A$ and ultimately return to $A$ without reaching $B$ (see
Fig.~\ref{fig:sketch}).

In FFS, we use a series of interfaces to sample the phase space
between $A$ and $B$ in stages.  At each stage, an ensemble of paths is
generated by firing off trajectories from points on an interface that
have been obtained in the previous stage; each of these trajectories
is terminated as soon as it reaches either the next interface or
$\lambda_0$ (see section \ref{sec:FFS}). We denote the average time spent at  order parameter $q$, for a trial run that is  fired from interface $\lambda_i$ (and
terminated at $\lambda_{i+1}$ or $\lambda_0$) in the FFS procedure, by
$\pi_+(q;\lambda_i)$. As shown in the appendix, 
 $\tau_+(q;\lambda_0)$ is then given by:
\begin{equation}
\tau_+(q;\lambda_0) = \pi_+(q;\lambda_0) + \sum_{i=1}^{n-1} \pi_+(q;\lambda_i) \prod_{j=0}^{i-1} P(\lambda_{j+1} | \lambda_j).
\label{eq:p_q_j}
\end{equation}

The factor $\prod_{j=0}^{i-1} 
P(\lambda_{j+1}|\lambda_j)$ reweights the distribution $\pi(q;\lambda_i)$ to correct for the enhanced sampling at interface $i$ which has been achieved by the FFS procedure. This factor is a direct output
from the FFS simulation (see section \ref{sec:FFS}, Eq.~\ref{eq:kAB}, and the appendix). The FFS calculation for the forward transition thus yields $k_{AB}$,
$\Phi_A$ and $\tau_+(q;\lambda_0)$.

To calculate $\rho(q)$, we also need to evaluate $\psi_B(q)$ in Eq.~\ref{eq:rho_phi_A_B_q}, by carrying out an FFS calculation in the reverse direction, from $B$ to $A$. The entire FFS algorithm is carried out in reverse: in the initial, brute-force simulation, we begin with the system in state B and collect crossings of interface $\lambda_n$ coming from $B$. We fire trajectories from $\lambda_i$ which either reach $\lambda_{i-1}$ or return to $\lambda_n$. The result is a value for the reverse rate constant $k_{BA}$, the flux $\Phi_B$ and the distribution functions $\pi_-(q;\lambda_i)$ for the order parameter $q$, sampled over the ensemble of paths that are fired from interface  $\lambda_i$ and terminated at $\lambda_{i-1}$ or $\lambda_n$ in the reverse FFS procedure. These are related to  the distribution function $\tau_-(q;\lambda_n)$ for all trajectories leaving $\lambda_n$ from $B$ by:
\begin{equation}
\tau_-(q;\lambda_n) = \pi_-(q;\lambda_n) +  \sum_{i=n-1}^{1} \pi_-(q;\lambda_i) \prod_{j=n}^{i+1} P(\lambda_{j-1} | \lambda_j),
\label{eq:p_q_jrev}
\end{equation}
where $P(\lambda_{j-1}|\lambda_j)$ are the conditional probabilities
of reaching interface $j-1$ from $\lambda_j$, evaluated in the reverse
FFS procedure. The distribution $\psi_B(q)$ is then given by:
\begin{eqnarray}
\psi_B(q)&=& p_B \Phi_B \, \tau_-(q;\lambda_n).
\label{eq:phiB_q}
\end{eqnarray}

To obtain $p_A$ and $p_B$ in Eqs(\ref{eq:phiA_q}) and (\ref{eq:phiB_q}), we note that in steady state 
\begin{equation}
\label{prob1}
p_A k_{AB} = p_B k_{BA},
\end{equation}
where $k_{AB}$ and $k_{BA}$ are the forward and backward rate
constants measured in the forward and backward FFS calculations,
respectively. Since we are assuming a two state system
(i.e. ignoring intermediate states), we also know that $p_A
+p_B=1$. This implies that
\begin{equation}
\label{prob2}
p_A = \frac{k_{BA}/k_{AB}}{1+k_{BA}/k_{AB}}
\end{equation}
and
\begin{equation}
\label{prob3}
p_B = \frac{1}{1+k_{BA}/k_{AB}}.
\end{equation}

Combining all this information and using Eq(\ref{eq:rho_phi_A_B_q}), we can obtain the stationary distribution function $\rho(q)$ in the region $\lambda_0 < \lambda < \lambda_n$. This can be combined with brute-force sampling in the $A$ and $B$ basins to determine $\rho(q)$ over the full range of $q$ values, if required.

\section{Stationary distributions:  Method}
\label{sec:sm}
As discussed above, to obtain the stationary distribution $\rho(q)$ in the region $\lambda_0 < \lambda < \lambda_n$ we perform one FFS
simulation for the transition from $A$ to $B$ and one for the reverse
transition. For details on the implementation of the FFS method to compute the fluxes $\Phi_A$ and
$\Phi_B$, as well as the rate constants $k_{AB}$ and $k_{BA}$, we refer
to ref.~\cite{FFS}. Here, we briefly discuss how
$\tau_+(q;\lambda_0)$ and $\tau_-(q;\lambda_n)$ are obtained in practice. We
consider $\tau_+(q;\lambda_0)$; $\tau_-(q;\lambda_n)$ is obtained similarly, but in reverse, as described above. Our aim is to calculate the quantities $\pi_+(q;\lambda_i)$ and $P(\lambda_{j+1} | \lambda_j)$ in Eq.~\ref{eq:p_q_j} [or alternatively for the reverse transition, $\pi_-(q;\lambda_i)$ and $P(\lambda_{j-1} | \lambda_j)$ in Eq.~\ref{eq:p_q_jrev}]. Considering only the forward FFS procedure:  at each interface $\lambda_i$ we fire a total of $M_i$ trial
runs, each of which is terminated when the system reaches either 
$\lambda_{i+1}$ or $\lambda_0$. The probability
$P(\lambda_{i+1}|\lambda_i)$ is then estimated as
\begin{equation}
P(\lambda_{i+1}|\lambda_i) = \frac{N_i^{\rm s}}{M_i},
\end{equation}
where $N_i^{\rm s}$ is the number of trials that have successfully reached
$\lambda_{i+1}$. The function $\pi_+(q;\lambda_i)$ is given by
\begin{equation}
\pi_+(q;\lambda_i) = \frac{N_q}{\Delta q M_i},
\label{t_lambda}
\end{equation}
where $N_q$ is the number of times that during this set of trial runs
the order parameter of the system has a value between $q$ and
$q+\Delta q$. This is given by $N_q =  \Delta t \sum_{k=0}^{M_i}\sum_{s=0}^{n_k}
h_q(x_{k,s})$, where the double sum runs
over all the $n_k$ steps of all the $M_i$ trial paths starting at interface $\lambda_i$
and $h_q(x)$ is an indicator function that is one if during a time step
the system is between $q$ and $q+\Delta q$, and zero otherwise; again,
note that $n_k$ varies from one path to the next. The simulation timestep $\Delta t$ can in fact be neglected, since is it a constant and we plan to normalise $\rho(q)$ in any case. For
algorithms in which the time step can vary, $N_q$ is given by $N_q =
 \sum_{k=0}^{M_i}\sum_{s=0}^{n_k} \Delta t_{k,s} h_q(x_{k,s})$, where $\Delta t_{k,i}$ is the magnitude of 
time step $s$ of path $k$. To obtain $\tau_+(q;\lambda_0)$ we reweight $\pi_+(q;\lambda_i)$ and sum over all interfaces using Eq.~\ref{eq:p_q_j}. Once $\psi_A(q)$ and $\psi_B(q)$ have been obtained by performing FFS
simulations in both directions, $\rho(q)$ can be obtained via
Eq.~\ref{eq:rho_q}. We note that $\psi_A(q)$ and $\psi_B(q)$ should not be individually normalized, since they are not probability distribution functions in their own right, but simply contributions to the distribution function $\rho(q)$. If the average path length for paths originating in $A$ and $B$ is different, then the integral of  $\psi_A(q)$ and $\psi_B(q)$ over $q$ will be different. Normalizing $\psi_A(q)$ and $\psi_B(q)$ will result in incorrect relative contributions to $\rho(q)$ from trajectories originating in $A$ and in $B$. We also note that, when evaluating $N_q$, it is important not to double-count the start and end points of trial runs - if the initial point of a trial run is deemed to count towards the $N_q$ histogram for that interface, then the final point should not count as it will be counted as an initial point in the histogram for the next interface.

The above procedure generates $\rho(q)$ in the region $\lambda_0 < \lambda < \lambda_n$. To obtain the full distribution  $\rho(q)$, we sample using conventional, brute-force simulation the steady-state distribution for the order parameter $q$ in the $A$ and $B$ regions. This will result in distributions for $\lambda < \lambda_0 + \Delta \lambda$ ($A$ region) and $\lambda > \lambda_n - \Delta \lambda$ ($B$ region), where $\Delta \lambda$ is a small overlap. An easy way to fit these curves together is to take their logarithms: the three overlapping parts for $\log \rho(q)$ can then be fitted together by a least squares fitting procedure (since a constant may be added to each without affecting the distribution). The resulting full profile $\rho(q)$ is obtained by exponentiating $\log \rho(q)$, and the stationary probability distribution can finally be normalised.

\section{Stationary distributions of multiple order parameters}\label{sec:multop}

It is important to point out that
the procedure described in section \ref{sec:sm} may be adapted to allow the computation of stationary distribution functions of several order parameters (``free energy landscapes'' in the equilibrium case). In the case where we wish to find the stationary distribution (for $\lambda_0 < \lambda < \lambda_n$) as a function of two order parameters $q$ and $r$, Eq.~\ref{eq:rho_phi_A_B_q} is replaced by
\begin{equation}
\rho(q,r) = \psi_A (q,r) + \psi_B(q,r),
\label{eq:rho_phi_A_B_q_r}
\end{equation}
where
\begin{eqnarray}
\psi_A(q,r)&=& p_A \Phi_A \, \tau_+(q,r;\lambda_0)\\
\nonumber \psi_B(q,r)&=& p_B \Phi_B \, \tau_-(q,r;\lambda_n)
\label{eq:phiA_q_r}
\end{eqnarray}
and
\begin{eqnarray}
\tau_+(q,r;\lambda_0) &=& \pi_+(q,r;\lambda_0) + \sum_{i=1}^{n-1} \pi_+(q,r;\lambda_i) \prod_{j=0}^{i-1} P(\lambda_{j+1} | \lambda_j)\\\nonumber
\tau_-(q,r;\lambda_n) &=&  \pi_-(q,r;\lambda_n)  + \sum_{i=n-1}^{1} \pi_-(q,r;\lambda_i) \prod_{j=n}^{i+1} P(\lambda_{j-1} | \lambda_j).
\label{eq:p_q_j_r}
\end{eqnarray}
To evaluate the functions $\pi_+(q,r;\lambda_i)$ and $\pi_-(q,r;\lambda_i)$, we use a two-dimensional histogram $N_{qr}$ in the co-ordinates $q$ and $r$:
\begin{eqnarray}
\pi_+(q,r;\lambda_i) &=&  \frac{\Delta t N_{qr}}{\Delta q \Delta r M_i}
\end{eqnarray}
and the equivalent for the reverse FFS procedure. Here, $N_{qr}$ is the number of timesteps during the set of $M_i$ trial runs fired from $\lambda_i$ for which the system has a value of $q$ between $q$ and
$q+\Delta q$ and a value of $r$ between $r$ and $r+\Delta r$.

\section{Testing on a one-dimensional system}
\label{sec:oned}
As an initial test, we have applied the method to a single particle moving with Brownian dynamics 
in a one-dimensional double-well potential
\begin{equation}
V(x) = -bx^2 + cx^4,
\label{brown}
\end{equation}
with $b=2$ and $c=1$. Distance is measured here in units of $x_0$, while time is measured in units of $t_0$. The stationary distribution function, as a function of the x-co-ordinate, is the Boltzmann distribution:
\begin{equation}
\rho(x) \sim e^{-V(x)/k_BT}.
\end{equation}
The system is symmetric, so that that $p(A)=p(B)$. The particle moves according to:
\begin{equation}
v(t) = \frac{D}{k_BT}f(t) + \xi(t),
\end{equation}
where $f$ is the instantaneous force, $D$ is the diffusion constant and $\xi$ is chosen at random from a Gaussian distribution with zero mean and 
variance $\langle \xi^2 \rangle = 2D dt$ \cite{allentildesley}. We use the following values: 
$D = 0.01 x_0^2/t_0$, $k_BT=0.1$ and $dt=0.05 t_0$. We have carried out FFS simulations with  $n=8$ interfaces, $N_1=10000$ points at interface $\lambda_0$, and parameters as shown in Table \ref{maiertab1}. 
\begin{table}[h!] 
\centerline{
\begin{tabular}{||c|c|c||c|c|c||}
\hline {  } \,\, $i$ \,\, & $\lambda_i$ &   $M(\lambda_i)$  {  }  & \,\, $i$\,\, &  $\lambda_i$   &$M(\lambda_i)$ \\
\hline  0  &     -0.8                   &     100000             &     4     &  -0.1              &    25000 \\
\hline 1 &       -0.7                   &    250000          &    5     &  0.1                    &    12000    \\ 
\hline  2 &        -0.5                 &      17000          &   6     &    0.3                  &      10000 \\
\hline  3 &        -0.3                 &     70000          &   7     &0.5                       &         10000 \\
\hline
\end{tabular}}
\caption{Interfaces and the number of trials at each interface for the FFS sampling of the symmetric one dimensional double-well potential.}
\label{maiertab1}
\end{table}

\begin{figure}[h!]
\begin{center}
{\rotatebox{0}{{\includegraphics[scale=0.35,clip=true]{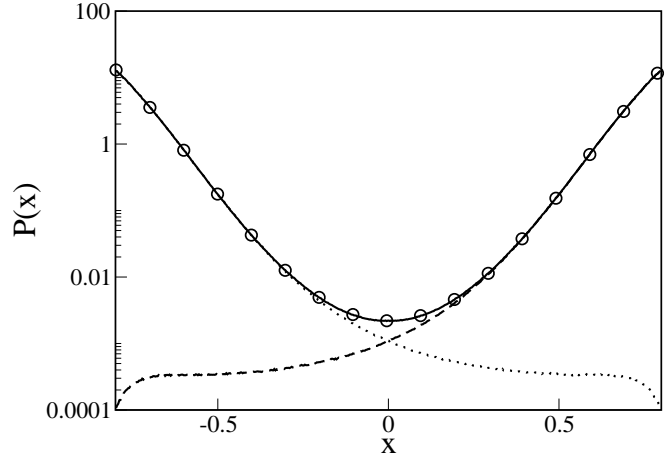}}}}
\caption{Stationary distribution (solid line) obtained using the procedure described above, compared to the 
normalized Boltzmann distribution (circles) for a symmetric double well potential. The dotted and dashed lines 
show $\psi_A(x)$ and $\psi_B(x)$ respectively.\label{fig1} }
\end{center}
\end{figure}

We obtained a forward rate constant $k_{AB} = 3.87 \pm 0.05 \times 10^{-6} t_0^{-1}$ (repeating twice to obtain error bar). Because of the symmetry of the problem, it was not necessary to carry out separate FFS calculations for the forward and backward transitions in this case - the backward probability distribution can be obtained from the forward one by a simple co-ordinate inversion. The stationary distribution obtained from the FFS calculation is compared to the expected Boltzmann distribution in Fig.~\ref{fig1}. 

We have also considered the asymmetric case, in which a term linear in $x$ is included in Eq.~\ref{brown}:
\begin{equation}
V(x) = ax + -bx^2 + cx^4,
\end{equation}
with $a=0.25$, $b=2$, $c=1$, $D = 0.01x_0^2/t_0$, $k_BT=0.1$ and $dt=0.05t_0$. In this case, $p(A) \ne p(B)$ and it is necessary to carry out FFS sampling in both directions. We carry out FFS simulations, again with $n=8$ interfaces and $N_1=10000$. For the forward transition, we used  $\lambda=x$, and for the backward transition, $\lambda=-x$. For both the forward and backward transitions, the parameters for the FFS runs were as shown in  Table \ref{maiertab2}.  
\begin{table}[h!]
\centerline{
\begin{tabular}{||c|c|c||c|c|c||}
\hline {  }  \,\, $i$ \,\, & $\lambda_i$ &   $M(\lambda_i)$  {  } & \,\, $i$ \,\,  & $\lambda_i$ &  $M(\lambda_i)$  \\
\hline 0  &      -0.8                    &       100000  &         4          &   -0.1           &       50000 \\
\hline  1  &     -0.7                    &     560000  &        5          & 0.1                 &       20000  \\
\hline  2  &      -0.5                   &      430000  &       6          &0.3                  &       12000  \\
\hline  3  &      -0.3                   &     170000   &      7          &0.5                   &        10000  \\
\hline
\end{tabular}}
\caption{Interfaces and the number of trials at each interface for the FFS sampling of the asymmetric one dimensional double-well potential.}
\label{maiertab2}
\end{table}

\begin{figure}[h!]
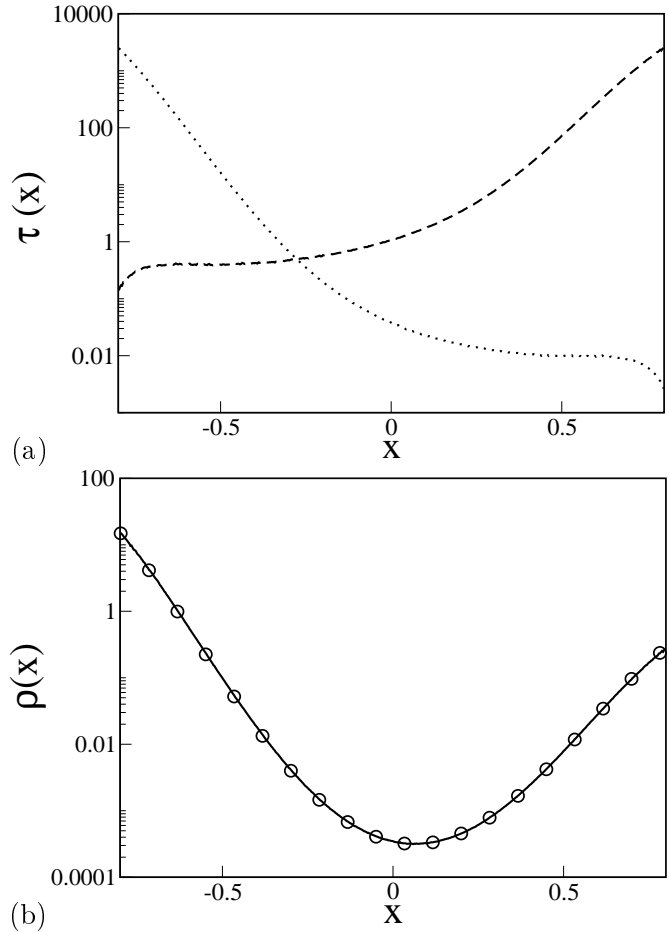

\begin{center}
\makebox[0pt][l]{(a)}{\rotatebox{0}{{\includegraphics[scale=0.35,clip=true]{raw_hists.eps}}}}\hspace{1cm}
\makebox[0pt][l]{(b)}{\rotatebox{0}{{\includegraphics[scale=0.35,clip=true]{final_hists.eps}}}}
\caption{ (a) Computed stationary distribution $\rho(x,y)$ as a function of $x$ for $y=-0.39$ (solid line), $y=-0.19$ (dashed line) 
and $y=0.01$ (dot-dashed line), compared to the expected Boltzmann distribution (indicated by circles) for the Maier-Stein system with $\epsilon=0.1$ and $\alpha=\mu=1$. (b) $\rho(x,y)$ as a function of $y$ for  $x=-0.312$ (solid line), $x=-0.152$ (dashed line) and $x=0.008$ (dot-dashed line).\label{fig2} }
\end{center}
\end{figure}

The forward and backward rate constants were calculated to be $k_{AB}=3.03 \pm 0.06 \times 10^{-7} t_0^{-1}$ and $k_{BA}= 3.96 \pm 0.03 \times 10^{-5}t_0^{-1}$, and the fluxes across the $A$ boundary were 
$\Phi_A=0.1526 \pm 0.0007 t_0^{-1}$ and $\Phi_B=0.3648 \pm 0.0001 t_0^{-1}$, respectively. Fig.~\ref{fig2}a shows $\tau_+(x;\lambda_0)$ and $\tau_-(x;\lambda_n)$, while Fig.~\ref{fig2}b shows $\rho(x)$, calculated from Eq.~\ref{eq:rho_q} and normalized. Excellent agreement is 
obtained with the expected Boltzmann distribution.

\section{Testing on the two-dimensional Maier-Stein system}
\label{sec:ms}
We now move on demonstrate the calculation of two-dimensional stationary distributions using a rare event problem in two dimensions that may be in or out of equilibrium - overdamped Brownian motion in the force field proposed by Maier and Stein \cite{maier1,maier2,maier3}:
\begin{eqnarray}\label{bd}
\dot{x} &=& f_x(\mathbf{x},t) + \xi_x(t)\\
\nonumber \dot{y} &=& f_y(\mathbf{x},t) + \xi_y(t),
\end{eqnarray}
where $\mathbf{x}=(x,y)$. The force field ${\bf{f}}=(f_x,f_y)$ (which is time-independent) is given by:
\begin{eqnarray}
f_x &=& x-x^3-\alpha x y^2\\
\nonumber f_y &=& -\mu y(1+x^2)
\end{eqnarray}
and the stochastic force ${\bf{\xi}}=(\xi_x,\xi_y)$ results from $\delta$-function-correlated white noise with variance $\epsilon$, such that
\begin{equation}
\langle \xi_i(t) \rangle = 0 \,\, ; \,\, \langle \xi_i(t) \xi_j(0)\rangle = \epsilon \, \delta_{ij}, \delta t, 
\end{equation}
where $i=x,y$. This system is bistable, with stable points at $(\pm 1,0)$ and a saddle point at $(0,0)$. 
When $\alpha = \mu$, the force field can be expressed as the gradient of a potential energy function and the system can be considered to be ``at equilibrium'', while 
when $\alpha \ne \mu$, the force field ${\bf{f}}$ cannot be expressed as the gradient of a potential and the system is thus intrinsically non-equilibrium. In these simulations, we use $\epsilon=0.1$. Taking $\lambda=x$, we follow the procedure described in section \ref{sec:multop} to calculate the stationary distribution for $-0.8 < x < 0.8$ as a function of the two order parameters $x$ and $y$. For the FFS calculations, we use 8 interfaces, $\lambda_0=-0.8$ and $\lambda_7=0.8$, and $N_1=100000$ initial configurations at $\lambda_0$. The parameters used are listed in Table \ref{maiertab3}.
\begin{table}[h!]
\centerline{
\begin{tabular}{||c|c|c||c|c|c||}
\hline \,\,$i$\,\, &  $\lambda_i$ &   $M(\lambda_i)$ & \,\,$i$\,\, & $\lambda_i$ &  $M(\lambda_i)$  \\
\hline 0 &       -0.              &  1000000      &     4        &0.0            &   200000           \\
\hline 1 &       -0.6             &   500000   &     5        & 0.2              &   120000   \\
\hline 2 &        -0.4            &   300000   &     6        &0.4               &   100000    \\
\hline 3 &        -0.2            &   250000   &     7        &0.6               & 100000  \\
\hline
\end{tabular}}
\caption{Interfaces and the number of trials per interface for the Maier-Stein system.}
\label{maiertab3}
\end{table}

\begin{figure}[h!]
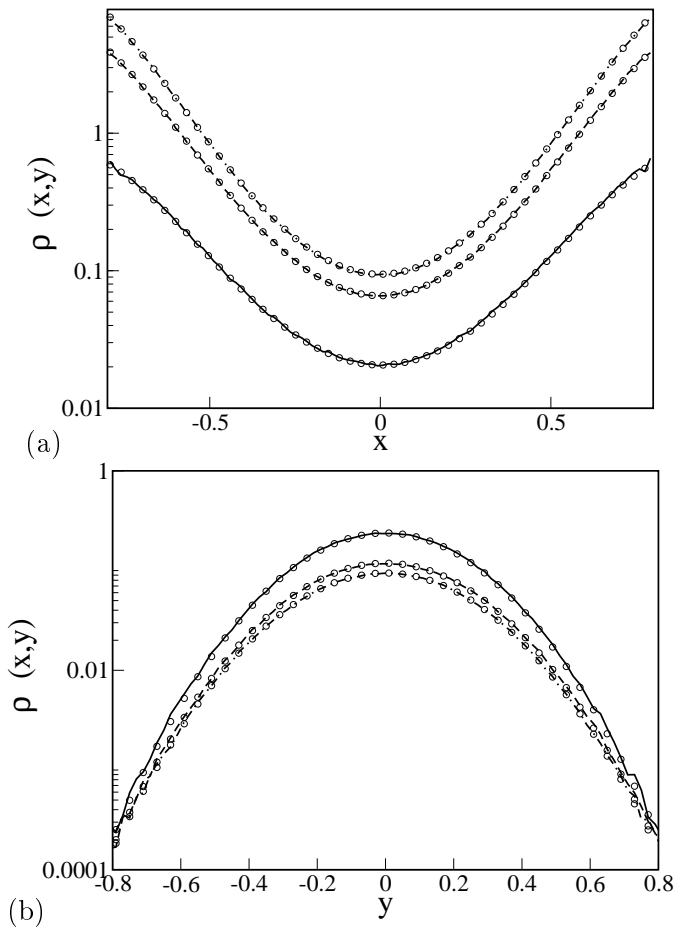

\begin{center}
\makebox[0pt][l]{(a)}{\rotatebox{0}{{\includegraphics[scale=0.35,clip=true]{Landscape_x.eps}}}}\hspace{1cm}
\makebox[0pt][l]{(b)}{\rotatebox{0}{{\includegraphics[scale=0.35,clip=true]{Landscape_y.eps}}}}
\caption{Asymmetric double well potential (a): $\tau_+(x;\lambda_0)/\Delta t$ (dotted line) 
and $\tau_-(x;\lambda_n)/\Delta t$ (dashed line)  (b): Final result for $\rho(x)$ obtained from Eq.~\ref{eq:rho_q} (solid line) 
compared to the expected Boltzmann distribution (circles).\label{fig3} }
\end{center}
\end{figure}

We initially consider an equilibrium case, with $\alpha=\mu=1$. In this case, the particle moves in the potential 
field $\phi(x,y) = \frac{y^2(1+x^2)}{2} - \frac{x^2}{2} + \frac{x^4}{4}$. Figures~\ref{fig3}(a) and (b) show the stationary distribution $\rho(x,y)$, as a function of $x$ for $y=-0.39$, $y=-0.19$  and $y=0.01$, and as a function of  $y$ for $x=-0.312$, $x=-0.152$ and $x=0.008$. In both panels, the results are in excellent agreement with the expected Boltzmann distribution (shown by circles).

\begin{figure}[h!]
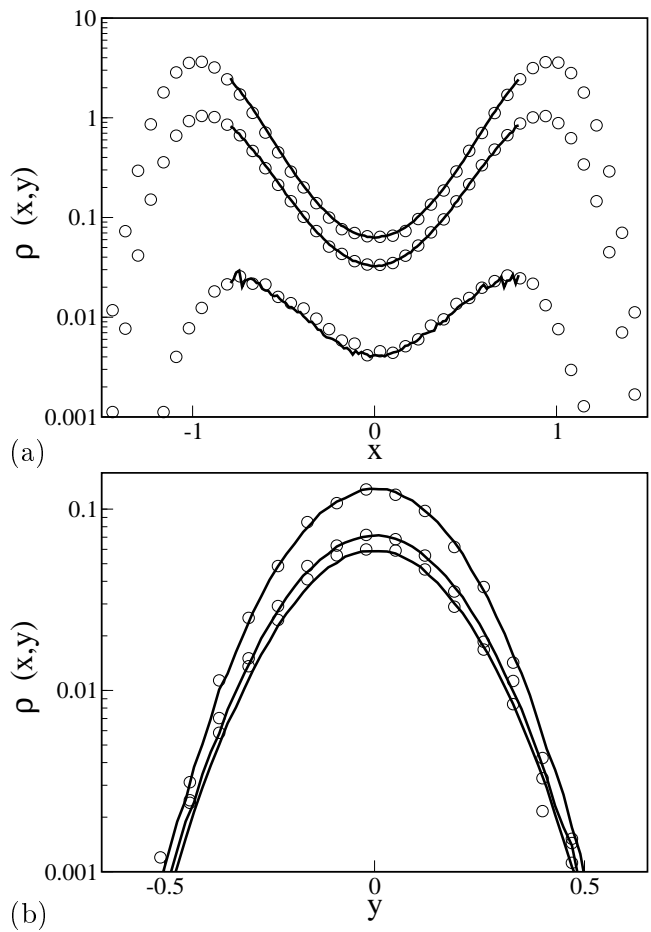

\begin{center}
\makebox[0pt][l]{(a)}{\rotatebox{0}{{\includegraphics[scale=0.35,clip=true]{Landscape_x_6.67_2.eps}}}}\hspace{1cm}
\makebox[0pt][l]{(b)}{\rotatebox{0}{{\includegraphics[scale=0.35,clip=true]{Landscape_y_6.67_2.eps}}}}
\caption{ $\rho(x,y)$ for the Maier-Stein system with $\alpha=6.67$, $\mu=2$ and $\epsilon=0.1$. (a): $\rho(x,y)$ 
as a function of $x$ for $y=-0.39$ (solid line), $y=-0.19$ (dashed line) and $y=0.01$ (dot-dashed line). (b):  $\rho(x,y)$ as a function of $y$ for $x=-0.312$ (solid line), $x=-0.152$ (dashed line) and $x=0.008$ (dot-dashed line). The results of long brute force simulations are indicated by circles.\label{fig4} }
\end{center} 
\end{figure}

We next discuss the non-equilibrium case ($\alpha \ne \mu$), taking
$\alpha=6.67$, $\mu=2.0$ and $\epsilon=0.1$. Fig.~\ref{fig4} shows
equivalent results to Fig.~\ref{fig3}, but this time the FFS results
are compared to stationary distributions computed from long
brute-force simulations. The brute-force simulation results are
normalised over all space; the FFS results are multiplied by a
constant scaling factor to bring them into agreement since they are
{\em{a priori}} normalised over the region $-0.8 < x < 0.8$ only. Very good agreement is observed.

\section{Homogeneous nucleation in a two dimensional Ising model
\label{sec:ising}}

We now address a rare event problem in a more complex system: homogeneous nucleation in a two-dimensional Ising model. For now, we confine ourselves to an equilibrium system without any external shear; non-equilibrium nucleation in an Ising model with an external shering field will be considered in future work \cite{future_paper}. The two-dimensional Ising model consists of an $L \times L$ square lattice of spins with nearest neighbour interactions and periodic
boundary conditions. Its Hamiltonian \cite{Binder}
\begin{equation}
H = - J \sum_{ij}^{\hspace{0.2cm}  '} \sigma_{i} \sigma_{j} - h \sum_{i} \sigma_{i},
\label{ham}
\end{equation}
where J is the coupling constant between neighboring spins
($\sigma_{i} = \pm 1$) and $h$ the external magnetic field.  The prime
indicates a sum over first nearest neighbour interactions only.  We simulate a system with $N =45 \times 45 = 2025$ spins, a positive magnetic field $\beta h=0.05$ and a positive
coupling constant $\beta J=0.65$, above the critical coupling
$J_c$. The thermodynamically stable state is therefore a
ferromagnetic one with net positive magnetization, meaning that the system tends to have the majority of its spins in  the ``up state''. However, the state with  an overall negative magnetization ({\em{i.e.}} spins
predominantly in the down state) is metastable and the system will remain in that state for a significant time if initialised with predominantly down spins. We aim to compute the free-energy barrier, as well as the rate constant, for transitions from the metastable ``down state'' to the thermodynamically stable ``up state''. We begin our simulations in the ``down state'' and  consider the formation of a
cluster of up spins, under conditions of moderate supersaturation (these conditions are identical to those used by Sear
\cite{sear}). All of our simulations are performed using a
Metropolis Monte Carlo algorithm, in which we attempt to flip each spin
once, on average, during each Monte Carlo cycle.

According to Classical Nucleation Theory \cite{Kelton}, the free energy cost of  forming a square nucleus of edge length $L$ is given by the sum of a line energy  and a surface energy:
\begin{equation}
\Delta G = 4\gamma L - 2hL^{2},
\label{barrCNT}
\end{equation}
where $\gamma$ is the interfacial free energy, $h$ is the driving force for nucleation (magnetic field), and $-2hL^{2}$  is the energy cost of flipping the whole square nucleus with area $L^{2}$.
Using Eq.~\ref{barrCNT}, the nucleation free energy barrier height is given by  
\begin{equation}
  \Delta G^{*} = \frac{2\gamma^{2}}{h}.
\label{topbarrCNT}
\end{equation}
Plugging in numbers, if we take the interfacial free energy to be $\beta \gamma = 0.74$
\cite{sear,onsager}, the barrier height as predicted by classical
nucleation theory is $\beta \Delta G^{*} \sim 22$. 

We have computed the nucleation free energy barrier using two simulation techniques:
umbrella sampling \cite{umbrella1,umbrella2,umbrella3,umbrella4} and FFS. In both cases, we characterize the extent of the transition using
a global order parameter, $q \equiv S$, the
total number of up spins in the system.  The free-energy barrier is
then defined as $\beta \Delta G(S) \equiv -\ln[\rho(S)/N]$, where $\rho(S)$ is
the probability of observing $S$ up spins in the stationary state. 

For our umbrella sampling calculations, we use a series of ``windows'', defined by a harmonic potential in
$S$, to bias the sampling of phase space
\cite{umbrella1,umbrella2,umbrella3,umbrella4}. We use 25 windows to cover the range $0 \le S \le 300$, with an overlap of 11 between neighbouring windows. We sample each window for 500000 MC cycles, and fit the resulting histograms together using a least-squares fitting procedure to obtain the free-energy profile in the range $0 \le S \le 300$. We do not attempt to calculate the barrier for values of $S$ greater than 300, since once the top of the barrier is crossed, the system is expected to evolve rapidly and we cannot reply on the assumption of local thermodynamic equilibrium. Moreoever, when $S$ is large, the growing nucleus is likely to interact with its periodic images in neighbouring cells, making the results highly system-size dependent. 

\begin{figure}[h!]
\begin{center}
{\rotatebox{0}{{\includegraphics[scale=0.35,clip=true]{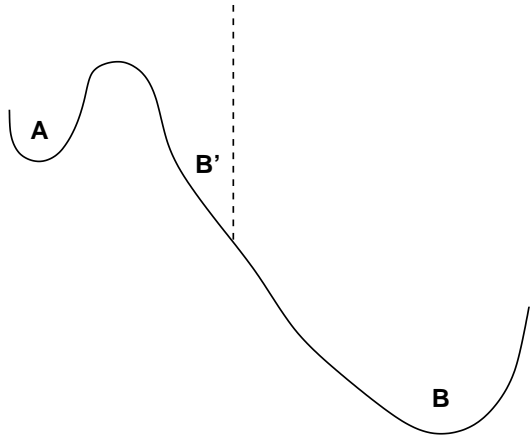}}}}
\caption{ A schematic view of the free energy landscape. A is the metastable ``down state'', B is the "real" thermodynamically stable 
state, and B$^{'}$ is the "artificial" stable state, constructed by introducing a reflecting wall at $S=S_{B^{'}}=1050$.
 \label{reflect}}
\end{center}
\end{figure}

The interfaces $\lambda_i$ for the FFS calculations are also defined
in terms of the order parameter $S$. To calculate the free-energy
barrier using FFS, we need to be able to sample the reverse
transition, from the thermodynamically stable ``up state'' to the
metastable ``down state''. In general, this is very difficult for a
nucleation problem, since the thermodynamic state is much more stable
than the metastable state and there is a very high free-energy barrier
for the system to return to the ``down state'', making the reverse
transition difficult to sample, even with FFS. We have overcome this
problem in this case by constructing a reflecting wall beyond the top
of the nucleation barrier. This wall is incorporated via a constraint
on the system dynamics: each trial move that leads to $S > S_{B'}$ is
simply rejected. Since we are only interested in the free-energy
profile in the region between $A$ and the top of the barrier, we may
perturb the free-energy landscape outside this region as we
choose. This fact, which is also exploited in umbrella sampling,
depends on the system being in equilibrium - for a driven system, we
would not be able to use this approach. The reflecting wall, located
at $S=S_{B^{'}}=1050$, replaces the $B$ state by an artificial stable
state $B^{'}$ (see Fig.~\ref{reflect}). 

The free-energy barrier for the
$B' \to A$ transition is much lower than that for the $B \to A$
transition, but the shape of the free-energy barrier on the $A$ side
remains unchanged. The use of the reflecting wall greatly facilitates
the FFS calculation for the reverse transition---it is
possible to carry out the reverse FFS calculations without the wall, but this is rather laborious as it requires a large
number of interfaces. We have verified that the location of the
reflecting wall is indeed well beyond the top of the free-energy
barrier, which is estimated to be at $200 < S < 280$.

State $A$ is defined by the first interface $\lambda_0=30$ - {\em{i.e.}} when $ 0 < S < 30$ the system is in the $A$ state. State $B'$ is defined by $\lambda_n=1000$ - {\em{i.e.}} when $ 1000 < S < 1050$ the system is in the $B'$ state. For our FFS calculations, we consider $N_1=50$ configurations at the first interface. The interfaces are located, both for the forward and backward sampling, at the values of $S$ given in table (\ref{maiertab4}), where we also list the  number of trials performed at each interface.
\begin{table}[h!]
\centerline{
\begin{tabular}{||c|c|c||c| c|c||c|c|c||}
\hline \,\,$i$\,\, &$\lambda_i$&$M_i$&\,\,$i$\,\,&$\lambda_i$&$M_i$&\,\,$i$\,\,&$\lambda_i$&$M_i$\\
\hline  0 &      30            & 1000 &  9       & 250       &  1000&   18           & 500 &  1000   \\
\hline  1&       50            & 1000 &  10       & 280      &  1000  &   19       &550                &  1000   \\
\hline 2 &       70            &  1000&  11       &300       &  1000 &   20     &600               &  1000 \\
\hline 3 &        100          &  1000&  12      &330        &  1000&   21    &650            &  1000\\
\hline 4 &        130          &  1000&  13      &350        &  1000 &  22    &700       &  1000\\
\hline 5  &        150         &  1000&    14   &380         &  1000&  23     &750           &  1000\\
\hline 6 &        180          &  1000&     15   &400        &  1000   &   24 &800          &  1000\\
\hline 7 &        200          &  1000&  16      &430        &  1000      &   25 &850            &  1000\\
\hline 8 &        230          &  1000&   17     &450        &  1000    &   26 &950                 &  1000\\
\hline
\end{tabular}}
\caption{Interfaces and the number of trials per interface for the FFS sampling for the two dimensional Ising nucleation problem.}
\label{maiertab4}
\end{table}

The FFS calculation for the forward transition from $A$ to  $B^{'}$ is 
straightforward. The flux $\Phi_A$ through $\lambda_0$ from $A$ is $\Phi_{A}=1.5 \times 10^{-5}$
MC step$^{-1}$ spin$^{-1}$ and the forward rate constant $k_{AB^{'}} = 2.8 \pm 0.3 \times
10^{-13}$ MC step$^{-1}$ spin$^{-1}$: this is in good
agreement with the value of  $3.3 \times 10^{-13}$
MC step$^{-1}$ spin$^{-1}$ computed for the same system by Sear {\em{et al.}}\cite{sear}. The computed forward rate constant does not depend on the reflecting wall position $S_{B'}$. This calculation also results in  the function $\tau_+(S;\lambda_0)$, as described in section \ref{sec:sm}. In the reverse direction, we use the same interfaces and sample from $\lambda_n$ to $\lambda_0$ as described in section \ref{sec:sm}. We obtain the flux $\Phi_B^{'} =1.4 \times 10^{-6}$ MC
step $^{-1}$ spin$^{-1}$ and the backward rate constant $k_{B^{'}A}= 2.0 \pm 0.2
\times 10^{-19}$ MC step$^{-1}$ spin$^{-1}$. In this procedure, we also compute the function $\tau_-(S;\lambda_n)$, as described in section \ref{sec:sm}. Combining the rate constants as in Eqs.(\ref{prob2}) and (\ref{prob3}), we obtain  $p_A= 7\times 10^{-7}$ and $p_{B'}=0.999$.
By means of Eqs.(\ref{eq:rho_phi_A_B_q}) and (\ref{eq:phiA_q}) we finally obtain
$\rho(S)$ for $30 < S < 1000$. Fitting this together with the
distribution obtained by conventional sampling in state $A$ (as
described in section \ref{sec:sm}), we obtain the free-energy
barrier. 
\begin{figure}[h!]
\begin{center}
\includegraphics[width=0.5\textwidth,clip=true]{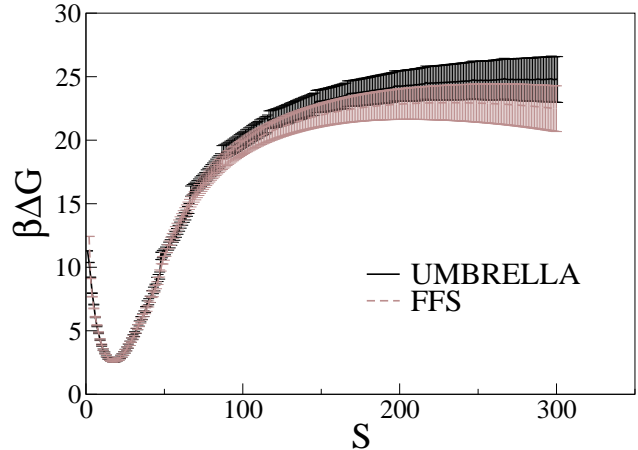}
\caption{ Free energy barrier for $\beta J =0.65$ and $\beta h=0.05$ calculated using FFS (dashed line; grey) and umbrella sampling (continuous line; black). Error bars are shown for both calculations.\label{barrier}}
\end{center}
\end{figure}

Figure~\ref{barrier} shows the results for the nucleation barrier,
$\beta \Delta G(S)$, in the range $0 < S < 300$, computed as
$-\ln\left[\rho(S)\right]$. The free-energy minimum at  $S \approx 20$
indicates that for this supersaturation, the system has a small number
of up spins even in the ``down'' state. The free-energy barriers, as
obtained by umbrella sampling and FFS, are $\Delta G^{\rm umbr} = 24.5 k_BT$ and
$\Delta G^{\rm FFS} = 23 k_B T$, respectively. These coincide within the error
bars, which for both schemes are on the order of $k_B T$. The computed
barrier heights also agree remarkably well with the CNT prediction
  of $22k_BT$.

\section{Genetic switch}
\label{sec:gs}

Our final test system is a biologically inspired non-equilibrium rare event problem: a model bistable genetic switch. This is a set of chemical reactions, representing protein-protein and protein-DNA interactions, as well as protein production and degradation, in a biological cell. The set of reactions shows two stable states, between which the system flips when simulated with stochastic dynamics. This is a particular case of the ``exclusive'' bistable
genetic switch studied by Warren et al. \cite{Warren05}. The system
does not obey detailed balance, and is therefore out of equilibrium. The set of chemical reactions which we simulate is given in scheme (\ref{eq:original_set}). 
\begin{subequations}
\label{eq:original_set}
\begin{align}
&\mathrm{Reaction} & \quad &\mathrm{Rate} & \quad & \mathrm{Reaction} & \quad & \mathrm{Rate} \nonumber\\
\hline
&\mathrm{A}+\mathrm{A} \rightleftharpoons \mathrm{A}_2 & \quad & k_{\rm f},\:k_{\rm b}
& \quad &\mathrm{B}+\mathrm{B} \rightleftharpoons \mathrm{B}_2 & \quad & k_{\rm f},\:k_{\rm b} \\
&\mathrm{O} + \mathrm{A}_2 \rightleftharpoons \mathrm{O}\mathrm{A}_2 & \quad & k_{\rm on},\:k_{\rm off}
& \quad & \mathrm{O} + \mathrm{B}_2 \rightleftharpoons \mathrm{O}\mathrm{B}_2 & \quad & k_{\rm on},\:k_{\rm off} \\
&\mathrm{O} \to \mathrm{O} + \mathrm{A} & \quad & k_{\rm prod} & \quad & \mathrm{O} \to \mathrm{O} + \mathrm{B} 
& \quad & k_{\rm prod}\\
&\mathrm{O}\mathrm{A}_2 \to \mathrm{O}\mathrm{A}_2 + \mathrm{A} & \quad & k_{\rm prod} & \quad &
\mathrm{O}\mathrm{B}_2 \to \mathrm{O}\mathrm{B}_2 + \mathrm{B} & \quad & k_{\rm prod} \\
&\mathrm{A} \to \emptyset & \quad & \mu & \quad & \mathrm{B} \to \emptyset & \quad & \mu
\end{align}
\end{subequations}

\begin{figure}
\begin{center}
\includegraphics[width=0.5\textwidth]{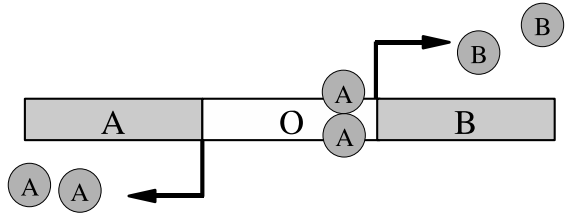}
\caption{Schematic representation of our model switch, corresponding to Eq.~\ref{eq:original_set}. Two divergently-transcribed genes
are under the control of a shared regulatory binding site on the DNA (the operator). Each protein can bind, in homodimer form, to
the operator and block the production of the other species.}
\label{fig:diagram}
\end{center}
\end{figure}

Our model switch is shown schematically in Fig.~\ref{fig:diagram}. 

It consists of two genes, which encode proteins A and B. Proteins A and B can form homodimers ${\rm{A_2}}$ and ${\rm{B_2}}$, as in Eq.(\ref{eq:original_set}a). The production rates for A and B depend on the state of the DNA sequence O, which is a regulatory site to which either ${\rm{A_2}}$ or ${\rm{B_2}}$ can bind. When O is free (not bound by either dimer), both genes can randomly be activated and produce either protein A or B with the same production rate $k_{\rm prod}$, as in Eq.(\ref{eq:original_set}c). When an ${\rm{A_2}}$ dimer is bound to O (Eq.(\ref{eq:original_set}b)), the production of B
is blocked. Conversely, when a ${\rm{B_2}}$ dimer is bound to O (Eq.(\ref{eq:original_set}b)), the production of A
is blocked.  Both proteins can decay in the monomer form (accounting for active degradation processes and dilution in a growing 
cell), as in Eq.(\ref{eq:original_set}e). We assume that transcription, translation and protein folding can be modeled as a single Poisson process, representing protein production.
Clearly, when one species is abundant over the other one, many dimers of the majority species will be created, and the probability
of finding one of them bound to O will  be high. This effect will in turn lower the production rate of the minority species,
leading to a stabilization of the state. If a rare fluctuation, however, is able to build up a substantial number of the minority
species, these will in turn dimerize and bind to O, leading to a stochastic flip of the the switch.

A mean field analysis carried out in \cite{Warren05} confirms this intuitive fact analytically: 
for suitable choices of the reaction rates, the system exhibits three
fixed points: two symmetrical stable states, one rich in A and 
another rich in B, separated by one unstable state where the total
number of A equals the total number of B. The system can then be considered as a true bistable switch.

We have chosen parameters such that the system is bistable and symmetric. Using the production
rate $k_{\rm prod}^{-1}$ as a time unit, and indicating by $V$ the dimensionless volume
of the system, we use: $k_{\rm f}=5k_{\rm prod}V$, $k_{\rm b}=5k_{\rm prod}$ (so
that the equilibrium dissociation constant for dimerization is $K_{\rm D}^d=k_{\rm d}/k_{\rm f}=1/V$), $k_{\rm on}=5k_{\rm
prod}$, $k_{\rm off}=k_{\rm prod}$ (so that the equilibrium dissociation constant for operator binding is $K_{\rm D}^b=k_{\rm
off}/k_{\rm on}=1/(5V)$), $\mu=0.3k$. For simplicity, we will assume $V=1$. The system is simulated with an event-driven Kinetic Monte Carlo algorithm \cite{Gillespie77} which propagates the system according 
to the Chemical Master Equation, thus accounting for the stochasticity arising from
molecular discreteness and from the intrinsic randomness of reaction
events. The simulation variables are the numbers of molecules $n$ (copy numbers) of
each chemical species. Briefly, in this algorithm, one selects at each
simulation step a waiting time until the next reaction, and an
identity for the next reaction, from the correct probability
distributions. One then advances the simulation time by the chosen
waiting time, executes the chosen reaction, and updates the copy numbers of the species involved in the reaction. 

A natural ``order parameter'' for the system is the difference between the total
numbers of A and B proteins: $q=\lambda=n_{\rm A}+2n_{\rm
A_2}+2n_{\rm OA_2}-(n_{\rm B}+2n_{\rm B_2}+2n_{\rm OB_2})$. Since the system is symmetric, we know that $\Phi_{A}=\Phi_{B}$, $k_{AB}=k_{
BA}$, and therefore $p_{A}=p_{B}=0.5$. As this system is out of equilibrium, we do not sample a free-energy profile, but rather the non-equilibrium stationary probability distribution $\rho(q)=\rho(\lambda)$.

To measure the switching rate and $\rho(\lambda)$, we run an FFS simulation with 12 interfaces, setting $\lambda_0=-27,\lambda_n=27$, and using  $N_1=$10000 points at the first interface. The interfaces are positioned as shown in Table \ref{maiertab5}. 
\begin{table}[h!]
\centerline{
\begin{tabular}{||c|c|c||c|c|c||}
\hline \,\,$i$\,\, &  $\lambda_i$ &   $M_i$ & \,\,$i$\,\, & $\lambda_i$ &  $M_i$  \\
\hline 0 &        -27               &    50000  &    6      &  -8            &    250000\\
\hline 1 &       -25               &     50000 &     7     &   -5          &500000\\
\hline 2  &       -22              &       50000&    8     &  -2             &500000\\
\hline 3  &       -18              &    50000   &    9     &  0             &250000\\
\hline 4  &         -14          &100000       &     10    & 10            &50000\\
\hline 5  &         -12            &100000       &      11   & 20             &  50000\\
\hline
\end{tabular}}
\caption{Interfaces and the number of trials per interface for the FFS simulations for the model genetic switch.}
\label{maiertab5}
\end{table}

\begin{figure}
\begin{center}
\includegraphics[width=0.65\textwidth]{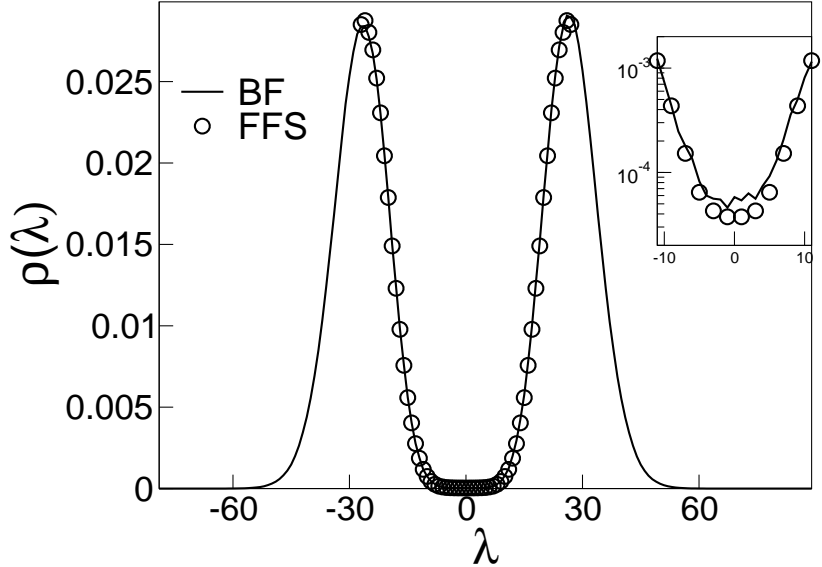}
\caption{Probability distribution as a function of the order parameter $\lambda$. The results are obtained both via long,
steady state simulations (continuous line) and Forward Flux Sampling (circles). The region around $|\lambda|=0$ can 
be accurately sampled only with FFS: the inset shows, on a  logarithmic scale, a much smoother profile of the region close to the
unstable steady state when FFS is used over Brute Force (BF). A scaling factor has been applied to the FFS results since they were originally normalised over $-27 < \lambda < 27$ while the BF results are normalised over $-\infty < \lambda < \infty$.
\label{fig:Eq}}
\end{center}
\end{figure}

We repeat the FFS sampling 10 times to obtain error bars. The result is  $k_{AB}=k_{BA}=(8.66\pm 0.07)\cdot 10^{-6}k_{\rm prod}^{-1}$. From the FFS calculations, we also obtain the function $\psi_A(\lambda)=\psi_A(q)$ as described in section \ref{sec:ms}, and since the system is symmetric, we can obtain  $\psi_B(\lambda)$ from $\psi_A(\lambda)$ by a simple inversion transformation. Combining $\psi_A(\lambda)$ and $\psi_B(\lambda)$, we arrive at $\rho(\lambda) = \rho(q)$ for $-27 < q < 27$, which is plotted in Fig.~\ref{fig:Eq} (a scaling factor is applied to account for the different normalisation to the brute force results). The distribution  is clearly bimodal and shows symmetric peaks whose positions correspond to the stable solutions 
of the mean field equations (\cite{Warren05}). As expected, a minimum in $\rho(\lambda)$ is observed for $\lambda=0$ (unstable solution of the mean field equations). 

 This system has a switching rate which is not exceedingly low, and we are also able to compute $\rho(q)$ using a brute force simulation of length $2\cdot 10^9k_{\rm prod}^{-1}$. The resulting stationary probability distribution is also shown in Fig.~\ref{fig:Eq}. Excellent agreement is obtained between the results of the FFS and brute force calculations. Because the system spends little time in the region  between the two basins, this part of $\rho(\lambda)$ is hard to calculate accurately with the brute force run. The inset in Fig.~\ref{fig:Eq} magnifies this region, showing the smooth profile produced by the FFS sampling.

 \section{Discussion}
 The key concept used here to obtain the stationary distribution in the unstable
 region between two stable states $A$ and $B$ is to add the
 contributions from the trajectories that start in $A$ and go to $B$
 or return to $A$, and those that start in $B$ and go to $A$ or return
 to $B$ (see Fig.~\ref{fig:sketch}). These contributions can be
 obtained by performing one FFS calculation starting in state $A$ and
 another starting in state $B$. For many rare event problems this is entirely possible - however, for systems where one state is very much more stable than the other, sampling the reverse transition ($B \to A$) may be computationally difficult, even with FFS. We have encountered this problem in the Ising nucleation example discussed here in section \ref{sec:ising}. For equilibrium systems, this problem can be overcome by imposing an artifical stable state, as demonstrated here for the case of
 nucleation. However, this trick is not applicable for
 non-equilibrium systems. In general, in equilibrium systems the flux
 between any two state points is zero in steady state, while for
 non-equilibrium systems this need not be the case. In these non-equilibrium systems,
 the stationary distribution depends upon the full history of the
 trajectories. This, in general, prohibits the introduction of
 artifical boundaries. In particular, while for equilibrium
 systems detailed balance and microscopic reversibility dictate that
 the forward and backward transition paths have to occupy the same
 region in state space, for systems that are out of equilibrium the
 backward and forward trajectories do not have to coincide; indeed, in
 these systems cycles in state space can occur. We have recently
 demonstrated that the switching pathways of genetic switches can follow
 such a scenario \cite{FFS}. If the forward and backward
 transition paths form a cycle in state space, then it is conceivable
 that the artifical stable state ``short cuts'' the cycle and generates a
 wrong ensemble of points from which trajectories are initiated in the reverse direction. It may be possible to devise alternative techniques for sampling the reverse transition in non-equilibrium systems, and this will be the subject of future work. 

 For the computation of free-energy barriers in equilibrium systems a
 wide range of numerical techniques is available \cite{daan}.
 The advantage of the scheme proposed here is that the free-energy can
 be directly obtained from an FFS simulation, obtaining simultaneously the rate
 constant, transition paths and free energy landscape. This is important because both the calculation of rate
 constants and the evaluation of free-energy barriers are
 computationally demanding, especially for large and complex systems.

 It has long been appreciated that free-energy barriers are critical
 quantities for understanding rare events in equilibrium systems, such
 as nucleation and protein folding. However, the ``barriers'', or minima 
in the stationary probabilities, that
 separate steady states in non-equilibrium systems are equally
 important, because the rate of switching from one steady state to the
 next is proportional to the probability of being at the top of the
``barrier'' \cite{Warren05}. Some such ``barriers'' have recently been determined experimentally, including bimodal distributions
 of protein concentrations for genetic switches like the one discussed
 in the previous section
 \cite{Gardner00,Ozbudak02}.  To our knowledge, this technique is the first to be proposed for efficient
 computation of stationary distributions for rare events in multi-dimensional
 non-equilibrium systems.  This should prove useful for enhancing our
 understanding of a range of important non-equilibrium rare event processes, as well as improving the efficiency of computation of free energy landscapes in equilibrium systems.

\section*{Acknowledgments}

The authors thank Vitaly Shneidman, Beate Schmittmann and Sorin T{\u{a}}nase-Nicola for their valuable advice.  Part of this work falls under the research program 
of the ``Stichting voor Fundamenteel Onderzoek der Materie (FOM)'', which is financially supported by the 
``Nederlandse Organisatie voor Wetenschappelijk Onderzoek (NWO)''. R.J.A. was funded by the European Union Marie Curie 
program and by the Royal Society of Edinburgh. 

\section*{Appendix}
In this appendix, we justify Eq.~\ref{eq:p_q_j}. Let us first imagine a very long brute force simulation trajectory which meanders around the basin of attraction of $A$, making occasional excursions towards $B$. We can divide each of these excursion into portions separated by successive crossings of interfaces $\lambda_0 \dots \lambda_n$. Consider the portion of an excursion between its leaving $A$ and either reaching $\lambda_1$ or returning to $A$. We denote the distribution function (averaged over many excursions) for points visited during this portion $\nu_0(q)$. Likewise, the distribution function (averaged over many excursions)  for points visited after crossing $\lambda_1$ and before reaching either $\lambda_2$ or $\lambda_0$ is denoted  $\nu_1(q)$, and we can also obtain distribution functions $\nu_i(q)$ for all interfaces $0 \le i < n$. It is important to note that the $\nu_i(q)$ are not normalised. In fact, the integral $\int dq \nu_i(q)$ contains information on the probability of an excursion reaching $\lambda_i$. Since our entire ensemble of excursions can be divided up in this way, we can write the entire distribution function $\tau_+(q;\lambda_0)$ as the sum of contributions from all the portions of trajectories: 
\begin{equation}
\tau_+(q;\lambda_0) = \sum_{i=0}^{n-1} \nu_i(q)
\end{equation}

Now let us consider the FFS procedure. Let us imagine we have generated a collection of points at interface $\lambda_i$. We fire $M_i$ trial runs from this collection of points and continue each one until either $\lambda_0$ or $\lambda_{i+1}$ is reached.  We plot a histogram $\pi_+(q;\lambda_i)$ of $q$ values for the points in this ensemble of trial runs. We have proved before \cite{FFS2} that the distribution of these trial paths is identical to the distribution of corresponding portions of the ``excursions'' from $A$ in a brute-force simulation, except that it is reweighted by a factor that depends on the probability of reaching $\lambda_i$ from $A$ - so that:
\begin{equation}\label{test1}
\pi_+(q;\lambda_i) = \frac{\nu_i(q)}{P(\lambda_i|\lambda_0)}  
\end{equation}
We have also proved before \cite{FFS2} that
\begin{equation}\label{test2}
P(\lambda_i|\lambda_0) = \prod_{j=0}^{i-1} P(\lambda_{j+1}|\lambda_j)
\end{equation}
for $i>0$ (for $i=0$, $P(\lambda_0|\lambda_0)=1$).

Combining (\ref{test1}) and (\ref{test2}), we arrive at
\begin{equation}\label{test3}
\pi_+(q;\lambda_i) = \frac{\nu_i(q)}{\prod_{j=0}^{i-1} P(\lambda_{j+1}|\lambda_j)}  
\end{equation}
for $i>0$ and $\pi_+(q;\lambda_0) = \nu_0(q)$. Rearranging Eq.~\ref{test3} and summing over interfaces, we arrive at 
\begin{equation}
\tau_+(q;\lambda_0) = \sum_{i=0}^{n-1} \nu_i(q) = \pi_+(q;\lambda_0) + \sum_{i=1}^{n-1}\pi_+(q;\lambda_i){\prod_{j=0}^{i-1} P(\lambda_{j+1}|\lambda_j)}  
\end{equation}
which corresponds to Eq.~\ref{eq:p_q_j}.


\bibliography{refs_barrier}

\end{document}